\documentstyle[12pt]{article}

  \font\msym=msym10
  \def\nn{\hbox{\msym N}}
  \def\rr{\hbox{\msym R}}

\scrollmode
\begin{document}

\vglue 2 true cm

\begin{center}
{\Large \bf The Periodic Table in Flatland}
\end{center}

\vspace{1cm}

\begin{center}

{\bf T.~NEGADI}

Laboratoire de Physique Th\'eorique,

Institut de Physique,

Universit\'e d'Oran,

Es-S\'enia, 31100 Oran, Algeria
\vspace{0.5cm}

{\bf M.~KIBLER}

Institut de Physique Nucl\'eaire de Lyon,

IN2P3-CNRS et Universit\'e Claude Bernard,

43 Boulevard du 11 Novembre 1918,

F-69622 Villeurbanne Cedex, France

\vspace{2cm}

{\bf Abstract}

\end{center}

\vspace{0.3cm}

The $D$-dimensional Coulomb system serves as a starting point for
generating generalized atomic shells. These shells are ordered
according to a generalized Madelung rule in $D$ dimensions. This
rule together with an {\it Aufbau Prinzip}
is applied  to produce a $D$-dimensional
periodic table. A model is developed to rationalize the ordering of the
shells predicted by the generalized Madelung rule. This model is based
on the introduction of an Hamiltonian,
invariant under the $q$-deformed algebra $U_q($so$(D))$,
that breaks down the SO($D+1$) dynamical symmetry of the
hydrogen atom in $D$ dimensions.
The $D=2$  case  (Flatland) is investigated with some
details. It is shown that the neutral atoms and the
(moderately) positive ions correspond to the values
$q=0.8$ and $q=1$, respectively, of the deformation
parameter $q$.

\vspace{3cm}

                 {\it Aufbau Prinzip}.

\newpage
\baselineskip 0.8 true cm

%===========par 1==================
\begin{center}
\section{Introduction}
\end{center}
% 1

There exits a huge number of articles and books devoted to
the classification of the chemical elements in the ordinary
space (with $D=3$ spatial dimensions). A nonexhaustive list
of works presenting quantitative achievements is provided
by Refs.~[1-26]. In recent years, the classification of molecules
in $D=3$ dimensions was also the subject of numerous studies
(see Refs.~[27-30] and references cited therein).

Group theoretical and Lie algebraic methods play an important role in
the classification of atoms and molecules. Along this vein, the authors
recently used the notion of $q$-deformed Lie algebra for the
classification of neutral atoms and positive ions in $D=3$ dimensions
\cite{NegKib}. From the mathematical point of view,
the approach in Ref.~\cite{NegKib} may be extended in
the case where $D$ is arbitrary.

The aim of this paper is two-fold: (i) to attack the problem of
the classification of chemical elements in a space-time with
$D$ spatial dimensions and (ii) to examine the special case where $D=2$.
Indeed, the idea of chemical elements in $D$ dimensions is already present
in the consideration of a $D$-dimensional hydrogen atom. During the last
four decades, hydrogen-like systems in $D$ dimensions were the subject of
numerous investigations (see for example Refs.~[31-39]).
As a next step, the study
of many-electron systems in $D$ dimensions is appealing. In this connection,
the formulation of the quantum-mechanical many-body problem in terms of
hyperspherical coordinates is of considerable interest [40-45]. The special
case $D=2$ is of great importance because of the specificities
that arise in this case. This explains why several works
have been devoted to Flatland \cite{Abo} as far as dynamical
systems and chemical elements are concerned [47-50].

The paper is organized as follows. In Section 2, we derive
in a compact way those aspects (energy and degeneracy) of
the $D$-dimensional
nonrelativistic
Coulomb system that are essential for
Sections 3 and 4. Sections 3 deals with an extension
of the Madelung rule and an {\it Aufbau Prinzip} in $D$ dimensions.
A model for the generalized $D$-dimensional Madelung rule is given
in Section 4. The case of Flatland ($D=2$) is discussed in Sections
3, 4, and 5.

\begin{center}
\section{The Hydrogen Atom in $D$ Dimensions}
\end{center}
% 2

{\it 2.1. The discrete spectrum}

The Schr\"odinger equation for a $D$-dimensional
hydrogen-like atom of nuclear charge $Ze$
                                   ($Ze > 0$) and
reduced mass $\mu$ reads
$$
- {\hbar^2\over 2\mu} \; \Delta \Psi + V \Psi \; = \;
  E \Psi,
\eqno (1)
$$
where $\Delta$ is the Laplace operator and $V$ the
potential energy in $D$ dimensions ($D \ge 2$).
We assume that
$$
V \; = \; - \; {Ze^2\over r},
\eqno (2)
$$
where $r$ is the hyperradius. Note that for $D=2$
and $D \ge 3$, the potential $V$ should be taken in the
form $V \sim   Ze^2 \ln r$ and
     $V \sim - Ze^2 r^{2-D}$, respectively, if one imposes that
$V$ satisfies the Poisson equation (cf.~Ref.~\cite{KveKat}).
Note also that the potential
energy $V$ could be chosen in the form (2) where $Z$
exhibits a dependence on the hyperspherical
angular coordinates
$\Omega \equiv (\theta_1, \theta_2, \cdots, \theta_{D-2}, \varphi)$
(see Refs.~[41,42]); the latter choice is especially appropriate
to take into account correlation effects in many-particle systems.
In the present paper, we take $V$ as given by (2) where $Z$ is a constant
for the sake of simplicity and in order to have
an expression that generalizes in a simple manner the usual
potential energy in $D=3$ dimensions. In view of the applications
to the construction of periodic tables, we are interested in the discrete
eigenvalues of the energy $E$ ($E < 0$) in Eq.~(1).
These eigenvalues can be found as
follows.

Let us look for a solution of Eqs.~(1-2) in the form
$$
\Psi (r, \Omega ) \; = \;
  R (r) Y_{[\ell]} (\Omega),
\eqno (3)
$$
where $Y_{[\ell]}$ denotes an hyperspherical
harmonic in $D$ dimensions. This leads, after
separation of variables, to the radial equation
$$
- \; {\hbar^2\over 2\mu} \;
  \bigg [ {d^2R\over dr^2} +
          {D-1\over r} \; {dR\over dr} \; - \;
          {\ell (\ell + D - 2)\over r^2} \; R \bigg ]
  - \; {Ze^2\over r} \; R \; = \; ER,
\eqno (4)
$$
where $\ell(\ell + D -2)$, with $\ell \in \nn$, is an eigenvalue
of the square of the angular momentum in $D$ dimensions
(see also the book by Avery \cite{Ave-liv}).
(In the $D=2$ case, the orbital angular
momentum $\ell$ is simply $|m_{\ell}|$.)
By putting
$$
\rho \; =
  \; 2 \, \alpha \, r,                          \qquad
  \alpha \; = \; {\sqrt {-2\mu E}\over \hbar }, \qquad
  k \; = \; {\mu Ze^2\over \alpha \hbar^2},     \qquad
  \nu \; = \; \ell + {1\over 2} \, (D-2)
\eqno (5)
$$
and by making the change of function
$$
R(r) \; = \;
  \rho^{- { (D-1)/2 } } \; W (\rho),
\eqno (6)
$$
Eq.~(4) can be transformed into the Whittaker differential equation
\cite{NikOuv}:
$$
W'' + \bigg (
  - {1\over 4} + {k \over \rho}
  + {{1\over 4} - \nu^2 \over \rho^2} \bigg ) W = 0.
\eqno (7)
$$
By retaining only the solutions regular at the origin
of Eq.~(7), we obtain the total wave function
$$
\Psi (r, \Omega) \; = \;
  (2 \, \alpha \, r)^\ell \, {\rm e}^{-\alpha r} \;
  _1F_1 \bigg ( {1\over 2} - k + \nu \; , \;
  2 \nu + 1 \; ; \; 2 \, \alpha \, r \bigg ) \; Y_{[\ell]} (\Omega),
\eqno (8{\rm a})
$$
where $_1F_1$ is the confluent hypergeometric function.

The discrete eigenvalues for the energy
$E$ are obtained by
requiring that the total wave function $\Psi$ belongs to
$L^2(\rr^D)$. This yields the quantization condition
$$
{1\over 2} - k + \nu \; = \;
  - n_r, \qquad n_r \in \nn,
% \hbox {\msym N},
\eqno (9)
$$
where $n_r$ is referred to as the radial quantum number.
By combining Eqs.~(5) and (9), we obtain that the discrete
energy spectrum is given by (cf.~Refs.~[31,32,36,38])
$$
E \; = \;
  {E_0 \over \big [ N + {1\over 2} (D - 1) \big ]^2}, \qquad
   E_0 \; = \; - \; {\mu (Ze^2)^2\over 2\hbar^2},        \qquad
     N \; = \; n_r + \ell,                               \qquad
     N \in \nn.
% \hbox {\msym N}.
\eqno (10)
$$
The possible values of $\ell$ are $\ell = 0, 1, \dots, N$.
(Observe that for $D=3$, we have $N=n-1$, where $n$ is the usual principal
quantum number. Note also that Eq.~(10) can
be applied to $D=1$ under the
condition to omit the state for which $N=0$.)
Equation (10) can be transformed to yield
$$
E \; = \; {E_0 \over \Lambda^2 + \big ( {D-1\over 2}\big )^2}, \qquad
  \Lambda^2 = \lambda (\lambda + D - 1),                       \qquad
  \lambda = n_r + \ell,
\eqno (11)
$$
an expression which constitutes a starting point for the developments
in Section 4. The quantity $\Lambda^2$ in Eq.~(11) turns out to be an
eigenvalue of the second-order Casimir operator of the
special orthogonal group SO($D+1$) in $D+1$ dimensions. In other words,
$\Lambda^2 \equiv \lambda (\lambda + D + 1 -2)$,
with $\lambda \in \nn$, is an eigenvalue of the square of the angular
momentum in $D+1$ dimensions. Equation (11) is the transcription in terms
of eigenvalues of the connection between the hydrogen-like atom in $D$
dimensions and the symmetrical spherical rotor in $D+1$ dimensions.
Moreover, the energy formula (11) reflects the SO($D+1$) dynamical
symmetry of the Coulomb system in $D$ dimensions.
This is reminiscent of
the application of the Fock stereographic
projection to the hydrogen atom in $D$ dimensions
(projection of $\rr^ D   $ onto the unit sphere of
               $\rr^{D+1}$).
As a point of fact, by applying the Fock transformation \cite{Foc} to the
$D$-dimensional Schr\"odinger equation (in momentum space)
for the hydrogen-like atom, we get [31,32,45]
$$
 {1 \over {\sqrt{- 2\, \mu \, E}}} \; {\mu Ze^2\over \hbar}
  \; = \;
  \lambda + {1\over 2} \; (D - 1), \qquad \lambda \in \nn,
\eqno (12)
$$
which leads to Eq.~(11).

Finally, by using Eq.~(9) the wave function (8a) becomes
$$
\Psi (r, \Omega) \; = \;
  N_{N,\ell} \, (2 \, \alpha \, r)^\ell \, {\rm e}^{-\alpha r} \;
  _1F_1 \bigg ( -n_r  \; , \;
  2 \nu + 1 \; ; \; 2 \, \alpha \, r \bigg ) \; Y_{[\ell]} (\Omega),
\eqno (8{\rm b})
$$
where the normalization constant  $N_{N,\ell}$  is
$$
N_{N, \ell} \; = \;
  {1 \over (2 \ell + D - 2) !} \;
  \bigg [ {(2\alpha)^D \; (N + \ell + D - 2)!\over
           (2N + D - 1) \; (N - \ell) !} \bigg ]^{1/2},
\eqno (13)
$$
up to an arbitrary phase factor. Equations (8b) and (13) are
in agreement with the results in Ref.~\cite{AveSomHen}.
Following the usual terminology used in
the $D=3$ case, we shall refer the wave function (8b) as an
atomic orbital. The various atomic orbitals corresponding to
given values of $N+1$ and $\ell$ define a shell
denoted as ($N+1, \ell$). As for the $D=3$ case, we shall use
the notation $\ell={\rm s}$, p, d, $\dots$
instead of   $\ell = 0$, 1, 2, $\cdots$ for the angular momentum
in ($N+1, \ell$).

{\it 2.2. Degeneracies of the spectrum}

The dimension of the subspace spanned by the angular wave functions
$Y_{[\ell]}(\Omega)$ corresponding to a given value of $\ell$ is
$$
g(\ell , D) \; = \;
  {(2\ell + D - 2) \; (\ell + D - 3 )!\over
   \ell ! \; (D - 2) !}, \qquad
  D \geq 3
\eqno (14{\rm a})
$$
that is nothing but the dimension of the irreducible representation
$[\ell, 0, 0, \cdots, 0]$ (with $D-2$ entries) of the geometrical
symmetry group SO($D$) of the hydrogen-like atom in $D$ dimensions.
Indeed, $g(\ell,D)$ corresponds to the essential
(i.e., geometrical or orbital) degeneracy of the
energy level
$E$ given by Eq.~(10). The dimension formula
(14a) can be simply obtained \cite{BacRic}
(see also [33,45])
as the difference between the
dimension of the space of the homogeneous polynomials with degree $\ell$ in
$\rr^D$ and the dimension of the space of the homogeneous polynomials
with
degree $\ell- 2$ in $\rr^D$ (arising from the action of the
generalized Laplace operator on the homogeneous polynomials with degree
$\ell$ in $\rr^D$). As an example, Eq.~(14a) with $D=3$ gives back the
well-known result $g(\ell,3) = 2 \ell + 1$. For $D=2$, note that Eq.~(14a)
gives $g(\ell,2) = 2$ for $\ell \ne 0$ and is not valid for $\ell=0$.
(In fact, we have  $g(0,2) = 1$ that follows from $g(0,D) = 1$.)
Returning to the case where $D$ is arbitrary, we note that
$$
g(\ell, D) = g(\ell, D-1) + g(\ell-1, D)
\eqno (14{\rm b})
$$
so that $g(\ell,D)$ can be obtained in an iterative fashion
from $g(0,D) = 1$ and $g(\ell,2) = 2 - \delta({\ell,0})$.

We are now in a position to simply obtain the degree of
degeneracy of the energy level
$E$ given by Eqs.~(10-11). It is mentioned above that Eq.~(11)
for $E$ shows that the hydrogen-like atom in
$D$ dimensions is connected to the symmetrical spherical rotor in $D+1$
dimensions. Consequently, the degree of degeneracy of
% the energy level
$E$
% given by Eq.~(?)
is equal to the degree of degeneracy of the eigenvalue
$\lambda(\lambda + D + 1 - 2)$ of the second-order invariant
of SO($D+1$). The latter degree of degeneracy is given by
Eq.~(14a) with $\ell \mapsto N$ and $D \mapsto D+1$ since the relevant special
orthogonal group is SO($D+1$) and $\lambda = N = n_r + \ell$.
As a conclusion, the total degree of
degeneracy $g$ (covering essential and accidental degeneracies)
of the energy level
$E$ given by Eq.~(10) is
$$
g \equiv g (N, D + 1) \; = \;
  {(2N + D - 1) \; (N + D - 2)! \over
   N! \; (D - 1) !}, \qquad N \in \nn.
\eqno (15)
$$
The expression in the second right-hand side of Eq.~(15) is a
well-known result.
(See for instance Ref.~\cite{Ave-liv} where this expression
is derived from the theory of hyperspherical harmonics.)
The significance of Eq.~(15) is clear: for $N+1$ fixed, the $g$ orbitals (8b)
with $\ell = 0, 1, \cdots, N$ have the energy (10). By way of
illustration, we have $g = 2N+1$ for $D=2$ and
$g=(N+1)^2 = (n_r + \ell + 1)^2 = n^2$ for $D=3$. We thus recover the
Stoner number $g = n^2$ for the ordinary hydrogen atom.

\begin{center}
\section{Madelung Rule and Aufbau Prinzip}
\end{center}
% 3

We now examine the problem of the
distribution of the $Z$ electrons of a neutral atom on
the generalized atomic orbitals.
Such a distribution does not make sense if the energy of the shell
($N+1,\ell$) is given by Eqs.~(10-11) since the energy formula (10)
does not take into account the interelectronic repulsion between the $Z$
electrons of a complex atom. We shall introduce in Section 4 a symmetry
breaking mechanism in order to generate  monoelectronic  energy  levels
depending on $N+1$ and $\ell$. We thus continue with the filling of the
$Z$ electrons on atomic shells ($N+1,\ell$) that do not present anymore
accidental degeneracies.

In the ordinary case ($D=3$), it is a well-known fact that the
filling of the shells ($N+1,\ell$), with $N + 1 = n$, according
to the rule
$-$~$n$ increasing and, for $n$ fixed, to $\ell$ increasing~$-$
does not
correspond to a realistic situation for $Z \ge 19$
(see Ref.~\cite{Boh}). In the
$D=2$ case, it has been shown \cite{PyyZha},
on the basis of Hartree-Fock calculations,
that this rule ($N+1$ increasing and, for $N+1$ fixed, $\ell = |m_{\ell}|$
increasing) stops working for $Z \ge 15$:
for $Z=15$, the energy of the shell (4,s) is lower
          than the one of the shell (3,d). In both cases, an
alternative rule for the filling of the shells has to be found.
When $D=3$,
such a rule is known as the Madelung (or Madelung-Klechkovskii, or
Madelung-Goudsmit-Bose) rule [2-5]. (In the $D=3$ case, this rule
was implicit in the 1922 Bohr {\it Aufbau Prinzip} \cite{Boh}
and was set out in  1926 by Madelung \cite{GouRic}.)
We shall formulate this rule in the case where $D$ is arbitrary ($D \ge 2$)
and we shall apply it to the particular cases $D=2$ and 3.

The generalized
Madelung rule can be expressed in the following way. The electronic shells
($N+1, \ell$) are filled according to: (i) $N+1+\ell$ increasing and (ii)
$N+1$ increasing for $N+1+\ell$ fixed. The obtained
filling can be described by the following sequence of shells
$$
  (1,{\rm s}) < (2,{\rm s}) < (2,{\rm p})
< (3,{\rm s}) < (3,{\rm p}) < (4,{\rm s})
< (3,{\rm d}) < (4,{\rm p}) < (5,{\rm s}) <
$$
$$
  (4,{\rm d}) < (5,{\rm p}) < (6,{\rm s})
< (4,{\rm f}) < (5,{\rm d}) < (6,{\rm p})
< (7,{\rm s}) < (5,{\rm f}) < (6,{\rm d})
< (7,{\rm p}) < \cdots
\eqno (16)
$$
The ordering (16) is identical to the one predicted
by the usual Madelung rule corresponding to $D=3$.

We are now prepared for describing an {\it Aufbau Prinzip}
in $D$ dimensions. Each $(N+1, \ell)$
shell in (16) can be filled with $2 g(\ell, D)$ electrons,
where $g(\ell, D)$ is given by (14a),
the factor 2 coming from the fact that each
% ($N+1, \ell$)
orbital gives rise to two spin-orbitals. [As for $D=3$, we admit
that the spectral group that labels the spin is SU(2) and
that the Pauli principle works.]
The generalized Madelung rule can be depicted by the diagram
   \[
   \begin{array}{l}
    {[1,1]} \\
    {[2,2] \; [3,2]} \\
    {[3,3] \; [4,3] \; [5,3]} \\
    {[4,4] \; [5,4] \; [6,4] \; [7,4]} \\
    {[5,5] \; [6,5] \; [7,5] \; [8,5] \; [9,5] \cdots} \\
    {\cdots}
   \end{array}
   \]
$$
\eqno (17)
$$
In (17) the rows are labelled with $N+1 = 1, 2, 3, \cdots$ and the
  columns with                     $\ell= 0, 1, 2, \cdots$
and the entry at the intersection of the $(N+1)$th row and $\ell$th
column is denoted as $[N+1+\ell, N+1]$
(with, for fixed $N+1$, $\ell = 0, 1, \cdots, N$).
This diagram provides us with the skeleton of a periodic table where the
places for the
various ``chemical elements'' are obtained by replacing each entry
$[N+1+\ell, N+1]$ of the diagram by a block $\{ \cdots \}$. Each block
corresponds to given values of $N+1$ and $\ell$. The block $\{ \cdots \}$
associated to the entry
$[N+1+\ell, N+1]$ of the diagram contains
$2 g(\ell, D)$ elements.

In the
$D=3$ case, this {\it Aufbau Prinzip}
leads to the periodic table \cite{Kib89}
   \[
   \begin{array}{l}
   \{ 001, 002 \} \\
   \{ 003, 004 \} \; \{ 005, 006,\cdots , 010 \} \\
   \{ 011, 012 \} \; \{ 013, 014,\cdots, 018 \} \;
   \{ 021, 022, \cdots , 030 \} \\
   \{ 019, 020 \} \; \{ 031, 032, \cdots , 036 \} \;
   \{ 039, 040, \cdots , 048 \} \;
   \{ 057, 058, \cdots , 070 \} \\
   \{ 037, 038 \} \; \{ 049, 050, \cdots , 054 \} \;
   \{ 071, 072, \cdots , 080 \} \;
   \{ 089, 090, \cdots , 102 \} \; \{ \cdots \\
   \{ 055, 056 \} \; \{ 081, 082, \cdots , 086 \} \;
   \{ 103, 104, \cdots , 112 \} \;
   \{ \cdots \\
   \{ 087, 088 \} \; \{ 113, 114, \cdots , 118 \} \;
   \{ \cdots \\
   \{ 119, 120 \} \; \{ \cdots \\
   \{ \cdots
   \end{array}
   \]
$$
\eqno (18)
$$
Each neutral atom in (18) is characterized by its atomic number
$Z$ ($001 \equiv 1$,
     $002 \equiv 2$,
     $003 \equiv 3$, $\cdots$).
In the table (18), the various periods of the
Mendeleev table are identified by reading the table in the
dictionary order,
as in the diagram (17),
and by stopping at the end of a ($N+1$,p) shell (except for the first period).
It is to be reminded that the table (18) can be rationalized
through the use of the group SO($4,2$)$\otimes$SU(2).
(For more details, see Refs.~[10,12,16,23].)

   In the $D=2$ case, the generalized Madelung rule gives also the ordering
described by the sequence (16) with $(N+1, \ell) \equiv (N+1, |m_{\ell}|)$.
The filling,
with two [$2 g(         0,2)=2$] electrons
in each $(N+1,          {\rm s})$ shell
and four [$2 g(\ell \ne 0,2)=4$] electrons
in each $(N+1, \ell \ne {\rm s})$ shell,
of the sequence (16) corresponds to the following periods
with their degeneracies (the degeneracy of a period is the
number of elements in the period)
 \[
 \begin{array}{l}
  (1,{\rm s})                                              :  2 \\
  (2,{\rm s}) \; (2,{\rm p})                               :  6 \\
  (3,{\rm s}) \; (3,{\rm p})                               :  6 \\
  (4,{\rm s}) \; (3,{\rm d}) \; (4,{\rm p})                : 10 \\
  (5,{\rm s}) \; (4,{\rm d}) \; (5,{\rm p})                : 10 \\
  (6,{\rm s}) \; (4,{\rm f}) \; (5,{\rm d}) \; (6,{\rm p}) : 14 \\
  (7,{\rm s}) \; (5,{\rm f}) \; (6,{\rm d}) \; (7,{\rm p}) : 14 \\
  \cdots
 \end{array}
 \]
$$
\eqno (19)
$$
The set of periods (19)
reflects (like in the  $D=3$  case) a period doubling, except for the
first period. The diagram (17) then leads to the following periodic table
\[
\begin{array}{l}
  \{ 001, 002 \} \\
  \{ 003, 004 \} \; \{ 005, 006, 007, 008 \} \\
  \{ 009, 010 \} \; \{ 011, 012, 013, 014 \} \; \{ 017, 018, 019, 020 \}
 \\
  \{ 015, 016 \} \; \{ 021, 022, 023, 024 \} \; \{ 027, 028, 029, 030 \}
 \; \{ 037, 038, 039, 040 \} \\
  \{ 025, 026 \} \; \{ 031, 032, 033, 034 \} \; \{ 041, 042, 043, 044 \} \;
    \{ 051, 052, 053, 054 \} \; \{ \cdots \\
  \{ 035, 036 \} \; \{ 045, 046, 047, 048 \} \; \{ 055, 056, 057, 058 \} \;
    \{ \cdots \\
  \{ 049, 050 \} \; \{ 059, 060, 061, 062 \} \; \{ \cdots \\
  \{ 063, 064 \} \; \{ \cdots \\
  \{ \cdots
\end{array}
\]
$$
\eqno (20)
$$
Here again,  the ``chemical elements''
in the blocks of (20) are characterized by their atomic number $Z$.
Furthermore, the elements in a given block
(corresponding to $[N+1+\ell, N+1]$)
can be distinguished by two heuristic quantum numbers, namely,
$m_{\ell} = \mp \ell$ and $\sigma = \mp 1/2$. Then, the address $Z$
of an element in the block attached to $[N+1, \ell]$ is given by
$$
  Z(N+1, \ell, m_\ell, \sigma ) =
  (N+1+\ell)^2 -
  {1\over 2} \; \bigg [ 1 + (-1)^{N+1+\ell}                 \bigg ]  -
              2 \bigg [ 2\ell + \delta (\ell+m_\ell ,0) - 1 \bigg ]  +
  \sigma +
  {1\over 2},
\eqno (21)
$$
with $N+1  = 1, 2, 3, \cdots   $;
     $\ell = 0, 1,    \cdots, N$;
 $m_{\ell} = \mp \ell          $; and
   $\sigma = \mp 1/2           $.
For instance,
$Z=11$       corresponds to $N+1=3$,
$\ell = 1$, $m_{\ell} = -1$, and $\sigma=-1/2$
while $Z=40$ corresponds to $N+1=4$,
$\ell = 3$, $m_{\ell} = +3$, and $\sigma=+1/2$.
In the block corresponding to
$[N+1+\ell, N+1]$, the number $Z$ increases
with $m_{\ell}$
($m_{\ell}$ passing from $-\ell$ to $+\ell$)
and, for $m_{\ell}$ fixed, with $\sigma$
($\sigma  $ passing from $-1/2 $ to $+1/2 $).

As for the $D=3$ case [17,20],
it is possible to define a key function for $D=2$.
The key function $Z_{\ell}$ is the atomic number $Z$ for which
(at least) one electron with angular momentum $\ell$ appears
for the first time.
(Note that this definition works for $D$ arbitrary too.)
By adapting to the $D=2$ case the derivation by Ess\'en \cite{Ess},
the key function reads
$$
Z_{\ell} \; = \;
2 \; \bigg [ \sum^{\ell - 1}_{N=0} \; 2 g (N,3)   \bigg ] + 1
\; = \;
2 \; \bigg [ \sum^{\ell - 1}_{N=0} \; 2 (2N + 1 ) \bigg ] + 1, \quad
{\rm for} \ \ell \ge 1,
\eqno (22)
$$
where the first factor 2 refers to the ``double shell''
structure
(that manifests itself in a diagram connecting the
hydrogen-like and Madelung orderings, see Ref.~\cite{Ess})
and the second to the spin. (Note that $Z_0=1$.)
A simple calculation gives
$$
  Z_\ell \; = \; 4\ell^2 + 1.
\eqno (23)
$$
For example, we have $Z_{\ell} = 1$, 5, 17, and 37 for
the first appearence of the s, p, d, and f electrons, respectively,
as can be checked from (20).

\begin{center}
\section{A Model for the Madelung Rule in Flatland}
\end{center}
% 4

As already mentioned at the beginning of
Section 3, the derivation of a model for justifying the
Madelung rule in $D$ dimensions cannot be carried out
with the help of the energy formula (11). Following Novaro
et al.~[9,11,13,14] and N\'egadi and Kibler \cite{NegKib},
a possible way to find an energy formula that
reproduces the generalized Madelung rule is to introduce,
in the context of $q$-deformations,  an anisotropic term
in the denominator of Eq.~(11). This can be done by replacing
in (11) the eigenvalue $\Lambda^2 \equiv \Lambda^2_{D+1}$ of the Casimir of the
group SO($D+1$) by $\Lambda^2_{D+1} + \alpha(q) \Lambda^2_{D}(q)$, where
$\Lambda^2_{D}(q)$ stands for an invariant operator of the $q$-deformed algebra
$ U_q ( {\rm so} (D) )$ and $\alpha(q)$ denotes a constant depending on the
deformation parameter $q$.
(See Refs.~[53-55]
for an introduction to $q$-deformed $-$~or quantum~$-$ algebras.)
To be more precise, the formula (11) is replaced by
$$
 E \; = \;
  {E_0 \over \Lambda^2_{D+1} + \alpha (q) \Lambda^2_D (q) +
             \big ( {D - 1 \over 2} \big )^2}
   \; = \;
  {E_0 \over \big ( N +  {D - 1 \over 2} \big )^2 + \alpha (q) \Lambda^2_D
(q)}.
\eqno (24)
$$
In terms of symmetries, the dynamical symmetry SO($D+1$) is replaced
by the symmetry  SO$(D+1) > U_q ( {\rm so} (D) )$,  where
the quantum algebra
$ U_q ( {\rm so} (D) )$ describes a symmetry breaking mechanism.

In the $D=3$ case, the formula (24) was applied in Ref.~\cite{NegKib} with
$$
  \Lambda^2_4     = (N + 1)^2 - 1 = n^2 - 1, \quad
  \Lambda^2_3(q)  = [ \ell ]_q [ \ell + 1 ]_q, \quad
  \alpha (q)      = 3 - {5\over 3} \; q,
\eqno (25)
$$
where the ($q$-deformed) quantities of type $[x]_q$ are defined by
$$
[ x ]_q \; = \; {q^x - q^{-x} \over q - q^{-1} }.
\eqno (26)
$$
The situation where $q=1$ gives back the Novaro  \cite{Nov73}  model for
neutral atoms
[corresponding to the ordering (16)]. Furthermore, it
was shown \cite{NegKib}
that Eq.~(24) with $D=3$ furnishes a model that describes
neutral atoms  for $q=0.9              $
[with the ordering                (16)] and
positive ions  for $1.15 \le q \le 1.30$
[with an  ordering different from (16)].

We now focus our attention on the $D=2$ case. In this case,
the dynamical symmetry SO(3) has to be replaced by the symmetry
$ {\rm SO}(3) \supset
  {\rm SO}(2) $ since the one-parameter Lie algebra
so(2) cannot be deformed. We can however consider a $q$-deformation of the
eigenvalue $\ell = |m_{\ell}|$ in order to mimick the approach followed
in Ref.~\cite{NegKib} for the $D=3$ case. Therefore we particularize, to the
$D=2$ case, Eq.~(24) in the form
$$
E \; = \; {E_0 \over
          N(N + 1 ) + \alpha (q)
\big ( [\ell]_q \big )^2 + {1\over 4} },
\eqno (27)
$$
where the anisotropy parameter $\alpha(q)$ is taken to be
$$
\alpha(q) = 3 - {5 \over 3} q,
\eqno (28)
$$
as for the $D=3$ case \cite{NegKib}. Then, Eq.~(27) becomes
$$
E \; = \; {E_0 \over
          \big ( N + {1\over 2} \big )^2 +
          \big ( 3 - {5\over 3}  \; q \big )
    \big ( [\ell ]_q \big )^2 },
\eqno (29)
$$
where the $q$-deformed number $[\ell]_q$ can be developed as
$$
[\ell]_q \; = \;
                 q^{  \ell - 1} +
                 q^{  \ell - 3} + \cdots +
                 q^{- \ell + 1} \quad {\hbox { for }} \; \ell \not = 0,
\eqno (30)
$$
an expression which may be prolonged by $[0]_q=0$. In order to find the
ordering of the shells $(N+1,\ell)$ afforded by the energy formula (29),
it is sufficient to consider the dimensionless quantity
$$
 \epsilon (N + 1, \ell ) \; = \;
     \sqrt {\bigg ( N + {1\over 2} \bigg )^2 +
\big ( 3 - {5 \over 3} q \big ) \big ( [\ell]_q \big )^2}
\quad {\rm with} \ N \in \nn \ {\rm and} \ \ell = 0, 1, \cdots, N
\eqno (31)
$$
that gives the energy position of the
shell $(N+1,\ell)$.

We have done an optimization of
Eq.~(31). As a result, for $q=0.8$ we find, with a reasonable
precision, that the energies given by Eq.~(29) reproduce the
ordering (16). The model inherent to Eq.~(29) is thus in
agreement with the two-dimensional
Madelung rule especially for the lower shells.
For instance, we have (for $q=0.8$) the sequence
$$
\epsilon (1,{\rm s})=0.5 \ < \
\epsilon (2,{\rm s})=1.5 \ < \
\epsilon (2,{\rm p})=2.0 \ < \
\epsilon (3,{\rm s})=2.5 \ < \
$$
$$
\epsilon (3,{\rm p})=2.8 \ < \
\epsilon (4,{\rm s})=3.5 \ < \
\epsilon (3,{\rm d})=3.6 \ < \
\epsilon (4,{\rm p})=3.7
\eqno (32)
$$
corresponding to $Z$ varying from 1 to 24.
The ordering (32) is the same as the one
found by Pyykk\"o and Zhao \cite{PyyZha} from Hartree-Fock
calculations using a Gaussian basis and assuming
$1/r$ Coulomb interactions.
We note that the fitting value $q=0.8$ is
of the same magnitude as the value $q=0.9$ obtained
in the $D=3$ case \cite{NegKib}.

In the situation where $q=1$, Eq.~(29) leads to the following
ordering
$$
  (1,{\rm s}) < (2,{\rm s}) < (2,{\rm p})
< (3,{\rm s}) < (3,{\rm p}) < (3,{\rm d})
< (4,{\rm s}) < (4,{\rm p}) < (4,{\rm d}) <
$$
$$
  (5,{\rm s}) < (5,{\rm p}) < (4,{\rm f})
< (5,{\rm d}) < (6,{\rm s}) < (6,{\rm p})
< (5,{\rm f}) < (6,{\rm d}) < (7,{\rm s})
< \cdots
\eqno (33)
$$
which is identical to the ordering for the positive ions
in the $D=3$ case [18,19,26]. Unfortunately, there does not exist
data (of  Hartree-Fock  type  for example) for the positive ions
in the $D=2$ case. Therefore, the sequence (33) should be considered
as a prediction for positive ions in two dimensions. To be more precise,
the passage from $q=0.8$ to $q=1$ for $D=2$ corresponds to the
passage from neutral atoms to moderately ionized atoms,
a situation that parallels the one for $D=3$ where $q$ goes from
0.9 (neutral atoms) to 1.15 (positive ions) \cite{NegKib}.
In the $D=2$ case, we can expect that the highly
ionized atoms correspond to increasing values of $q$ since the limiting
case of a hydrogen-like atom
(for which the energy $E$ increases with $N+1$)
is reached for $q=1.8$ [see Eq.~(29)].

\begin{center}
\section{Closing Remarks}
\end{center}
% 5

In the present paper, we concentrated on three points:
(i)   the extension of the Madelung rule in $D$ dimensions,
(ii)  an algorithm ({\it Aufbau Prinzip})
for constructing a periodic table in $D$
dimensions, and
(iii) a model for reproducing the ordering of the atomic shells in $D$
dimensions. For each point, a special emphasis was put on the $D=2$ case.

The periodic table for $D=2$, obtained by means of the Madelung rule,
resembles the one worked out in Ref.~\cite{Kib89} except that the blocks
contain two or four elements. We gave an address formula [Eq.~(21)]
for locating the element with atomic number $Z$ in the periodic table.
In addition, the key function for the first appearence
of an electron of type $\ell$ in the periodic table
was found to be $Z_{\ell} = 4 \ell^2 + 1$. This key function compares
           with $Z_{\ell} = 2 \ell^2 + 1$ that can be deduced from the work
by Asturias and Arag\'on \cite{AstAra},
where the Coulomb potential is taken in
a logarithmic form. (Note that the ``double shell'' structure mentioned
in Section 3 does not appear when the potential is logarithmic.)

The model developed in Section 4 on the
basis of Eq.~(24) relies on the connection
between the hydrogen atom in $D$ dimensions and the spherical rotor in $D+1$
dimensions. Two important ingredients were used in conjunction with the latter
connection, viz, the introduction of an anisotropic term as in the Novaro
\cite{Nov73}
model for $D=3$ and the introduction of a $q$-deformation as in the
N\'egadi-Kibler \cite{NegKib} model
for $D=3$. The model set up for $D$ arbitrary was applied to $D=2$ and it
was shown that the deformation parameter $q$ can discriminate between
neutral atoms ($q=0.8$) and
positive ions ($q=1$).

If we take the same notations for the chemical elements in two dimensions
as the ones employed in Ref.~\cite{PyyZha},
the periodic table (20) takes the form
\[
\begin{array}{llllllllllll}
 \{ {\rm H }^1 & {\rm He}^2 \} & & & & & & & & & & \\
               &               & & & & & & & & & & \\
 \{ {\rm Li}^3 & {\rm Be}^4    \} & &
 \{ {\rm B }^5 & {\rm N }^6 & {\rm F}^7 & {\rm Ne}^8 \} & & & & & \\
               &               & & & & & & & & & & \\
 \{ {\rm Na}^9 & {\rm Mg}^{10} \} & &
 \{ {\rm Al}^{11} & {\rm P}^{12}  & {\rm Cl}^{13} & {\rm Ar}^{14} \} & &
 \{ {\rm Sc}^{17} & {\rm Mg}^{18} & {\rm Cu}^{19} & {\rm Zn}^{20} \}  \\
               &               & & & & & & & & & & \\
 \{ {\rm K}^{15}  & {\rm Ca}^{16} \} & &
 \{ {\rm Ga}^{21} &  {\rm As}^{22} & {\rm Br}^{23} & {\rm Kr}^{24} \}
& & & & &
\end{array}
\]
$$
\eqno (34)
$$
if we restrict ourselves to $1 \le Z \le 24$. (The address $Z$ for each
element is indicated as a right upper subscript.) The table (34) is similar to
the table in Ref.~\cite{PyyZha}: the two tables exhibit the same blocks in
different arrangements. The main difference between the two tables is the
position of the {\it transition metals} (Sc, Mn, Cu, Zn).
The atomic configurations arising from the latter table
are in accordance with those derived
by Pyykk\"o and Zhao \cite{PyyZha}
from self-consistent field calculations, except
for the ``Scandium'' ($Z=17$). Indeed, according to our periodic table, the
element Sc has the configuration [Ar]4s$^2$3d, a fact to be contrasted
          with the configuration [Ar]4s3d$^2$ obtained
in Ref.~\cite{PyyZha}.

To close this paper, we would like to mention that the developments in the
present work could be considered as a first step towards a classification of
molecules in two dimensions. We hope to return on this subject in the future.

\bigskip

\begin{center}
{\bf Acknowledgements}
\end{center}

One of the authors (T.~N.) wishes to thank the
{\it Institut de Physique             de l'Universit\'e d'Oran} and the
{\it Institut de Physique Nucl\'eaire de l'Universit\'e Lyon~1}
for having made possible his stay in Lyon-Villeurbanne in December
1994.

\vfill\eject

% \begin{center}

\end{document}